\documentclass[pra,amssymb,superscriptaddress, longbibliography,twocolumn]{revtex4-2}

\usepackage{natbib}
\usepackage{amsmath}
\usepackage{amssymb}
\usepackage{multirow}
\usepackage{array}
\newcolumntype{P}[1]{>{\centering\arraybackslash}p{#1}}

\usepackage{mathtools}
\usepackage{mathrsfs}
\newcommand{\be}{\begin{equation}}
\newcommand{\ee}{\end{equation}}

\newcommand{\ra}{\rangle}
\newcommand{\la}{\langle}

\newcommand{\bea}{\begin{eqnarray}}
\newcommand{\eea}{\end{eqnarray}}

\usepackage{xcolor}
\usepackage{color}

\usepackage{hyperref}
\hypersetup{colorlinks=true,citecolor={blue},linkcolor={blue},urlcolor={blue}}
\begin{document}

% \title{Title 1: Two-qubits holonomic gates for quantum algorithms. \\
% Title 2:  Holonomic controlled quantum gates \\
% %and quantum universality\\ 
% Buzz words: holonomies, Thouless pumping, quantum computing, quantum universality, completeness }
% %Buzzwords: adiabatic, non-Abelian, Thouless pumping, quantum gates, }

\title{Holonomic multi-controlled gates for single-photon states}
%\title{Holonomic quantum gates for quantum algorithms}

\author{Carlo Danieli}
\email{carlo.danieli@cnr.it} 
\affiliation{Institute for Complex Systems, National Research Council (ISC-CNR), Via dei Taurini 19, 00185 Rome, Italy}%
\affiliation{Department of Physics and Astronomy, University of Florence, I-50019 Sesto Fiorentino, Italy}%
\author{Valentina Brosco}
\affiliation{Institute for Complex Systems, National Research Council (ISC-CNR), Via dei Taurini 19, 00185 Rome, Italy}%
\author{Claudio Conti}
\affiliation{Department of Physics, University of Sapienza, Piazzale Aldo Moro 5, 00185 Rome, Italy}%
\affiliation{Research Center Enrico Fermi, Via Panisperna 89a, 00184 Rome, Italy}%
\author{Laura Pilozzi}
\affiliation{Institute for Complex Systems, National Research Council (ISC-CNR), Via dei Taurini 19, 00185 Rome, Italy}%
\affiliation{Research Center Enrico Fermi, Via Panisperna 89a, 00184 Rome, Italy}%
%   efficient and experimentally realizable 

%
\begin{abstract}

Controlled and multi-controlled quantum gates, whose action on a target qubit depends on the state of multiple control qubits, represent a fundamental logical building block for complex quantum algorithms.
We propose a scheme for realizing this class of gates based on non-Abelian holonomies in modulated photonic waveguide networks. 
Our approach relies on linear photonic cicuits formed by two star networks coupled via a two-path circuit. A star network with $M$ peripheral waveguides coupled to a single central site, or $M$-pod, naturally generalizes the tripod structure used in non-Abelian Thouless pumping and stimulated Raman adiabatic passage (STIRAP). 
In the present work, we first analyze the minimal case of two connected tripods and design adiabatic driving loops that implement single-qubit, CNOT, and SWAP gates. We then show how  extending the approach to larger $M$-pod structures enables the realization of multiply controlled operations, which we exemplify by designing  Toffoli and the OR gate on two coupled pentapods. Finally, we demonstrate that networks of connected tripods can implement the Deutsch quantum query algorithm.

\end{abstract}

\date{\today}

\maketitle

\section{Introduction}

Non-Abelian holonomies~\cite{wilczek1984,anandan1988} represent the natural $U(N)$ generalization of Berry’s geometric phase~\cite{berry1984}. They were originally introduced by Wilczek and Zee in their seminal work~\cite{wilczek1984}, which elucidated the relation between  non-Abelian gauge theories and the geometric structure of Hilbert space emerging in the adiabatic evolution of quantum systems with degenerate energy levels.
Since then various experiments generated and detected non-Abelian holonomies in photonic platforms\cite{Sun2022,Chen2025}, acoustic waveguides \cite{oubu2022observation}, quantum optical setups \cite{zwanziger1990nonabelian,unanyan1999}, cold atoms \cite{sugawa2018}, and superconducting devices \cite{abdumalikov2013experimental}. 
A the same time, theoretical works highlighted their fundamental role in modeling quantum transport \cite{murakami2004,chen2019non,brosco2021nonabelian,Pilozzi2022} and describing topological phenomena \cite{taddia2017topological,danieli2025nonabelian,iadecola2016nonabelian}.

\begin{figure}[t!]
	\centering
    \includegraphics[width=0.925\columnwidth]{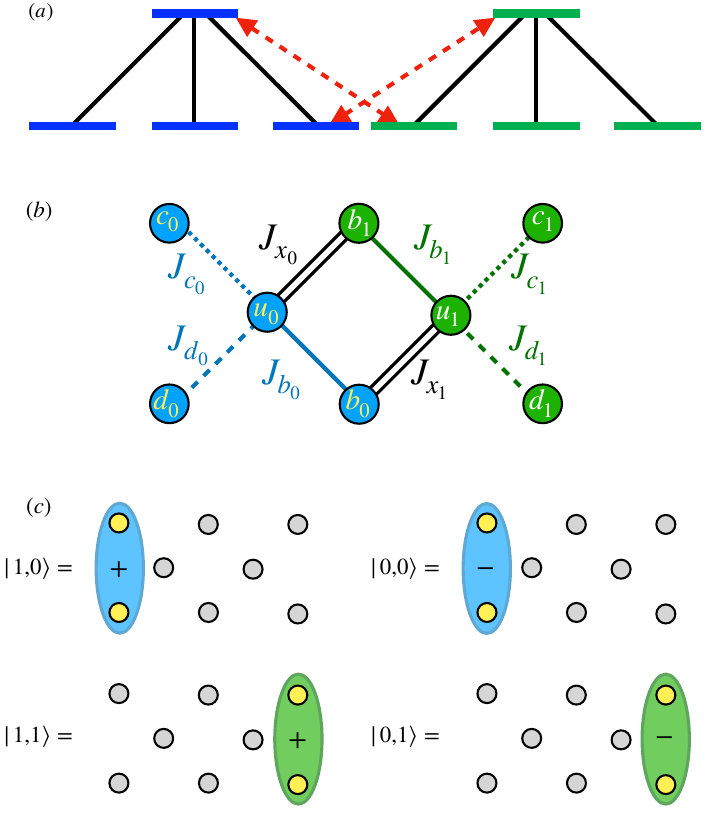}  
	\caption{
    (a) Scheme describing the structure of the Hamiltonian of two-coupled tripods. 
    (b) Schematic of the array composed of two tripods respectively colored in blue and green. The two tripods are connected via the hopping $J_{x_0}$ and $J_{x_1}$. 
	(c) Two qubits encoding states $|s,p\ra$ for the target qubit $p=0$ (upper row) and $p=1$ (lower row). At the initial parameters $J_{\ell_p} =0$ for $\ell = c,d,x$ and $J_{b_p}=J$ for $p=0,1$ of the driving cycle, the states correspond to the four zero-energy degenerate eigenstates.}
	\label{fig:1} 
\end{figure}

\noindent The basic relevance of non-Abelian holonomies is closely intertwined with their practical interest in quantum control \cite{duan2001geometric,brosco2008nonabelian,danieli2024parity} and quantum computing~\cite{ZANARDI199994,zhang2023a}.
%The idea of \textit{holonom}. 
In particular, as first demonstrated by Zanardi and Rasetti, holonomies can be exploited to implement a universal set of \textit{holonomic quantum gates} (HQGs)~\cite{ZANARDI199994}.
This idea, introduced in Ref.~\cite{ZANARDI199994}, is based on encoding qubits in the eigenstates of a degenerate Hamiltonian $H$ 
and performing unitary operations through the Wilczek–Zee holonomy by adiabatically modulating a set of classical control parameters defining $H$.
A key advantage of HQGs lies in their  robustness against  certain kinds of noise and decoherence~\cite{1997RuMaS..52.1191K,solinas2004robustness,sun2024suppression,sun2025decoherence}.
Since the pioneering work of Ref.~\cite{ZANARDI199994}, HQGs have been both theoretically proposed and experimentally realized across various physical platforms~\cite{Jones2000,Pachos2000optical,PhysRevLett.90.028301,Nagata2018,Bhattacharyya,chen2025realization}.
Furthermore, the potentially crucial role of HQGs in enabling more robust quantum computing architectures~\cite{david1985theory,548464,PhysRevA.57.127,BOYKIN2000101,paetznick2013universal}  has motivated various works aimed at realizing holonomic gates beyond the adiabatic regime~\cite{PhysRevLett.122.080501,zhao2023nonadiabatic,Neef:25,jin2025nonadiabatic,song2025}.

In both photonic and atomic implementations of holonomic systems, a crucial role is played by Hamiltonians with a tripod structure, such as those illustrated in Fig.~\ref{fig:1}(a).
This type of Hamiltonian exhibits two degenerate energy levels for any value of the coupling parameters, and the holonomy is typically generated by adiabatically controlling these couplings. 
Hamiltonians of this kind lie at the heart of non-Abelian Thouless pumping~\cite{brosco2021nonabelian} and stimulated Raman adiabatic passage (STIRAP) protocols~\cite{duan2001geometric,unanyan1999,zhang2023a,neef2023,song2025,pinske2020highly,pinske2022symmetry}.\\

\noindent In this work, we go beyond standard single-qubit holonomic protocols, by focusing on the design of 
multi-controlled holonomic gates with $m$ control qubits and one target qubit  within an adiabatically modulated array of $2(2^m+2)$  optical waveguides. 
This array consists of two coupled  $M$-pods, each having $M=2^{m}+1$ legs connected to a central waveguide. 
%two-qubit and multi-controlled holonomic gates within an adiabatically modulated array of $2^{2m+1}$  optical waveguides. 

\noindent The system proposed in the present work encodes  information in the state of a single-photon in a multimode setup and uses holonomies to manipulate this information implementing an holonomic version of linear optic quantum computing protocols\cite{babazadeh2017,lapkiewicz2011,kok2017}.\\ 
\noindent In the simplest case, the array structure is based on two tripods, $(m=1)$, connected {\sl via} hoppings that preserve the four-fold degeneracy of the single-photon space for any coupling strength as shown in Fig.~\ref{fig:1}(b).
By properly choosing the qubit encoding scheme and the driving cycles, we  engineer a universal set of holonomic transformations~\cite{vlasov2001clifford,tolar2018clifford,helsen2018representation,Bataille}.
In particular, beside single-qubit operations, such as Hadamard and  phase gates, we design controlled two-qubit gates, of the form $C_U=|0\ra \la0|\otimes \mathbb{I} + |1\ra \la1| \otimes U$ and the SWAP gate \cite{monroe1995demonstration}.

%We then consider larger structures featuring two coupled  $M$-pods each having $M=2^{m}+1$ legs and we discuss the implementation of multiply controlled gates. 
Multi-controlled quantum gates based on two coupled $M$-pods extend the ${\rm CNOT}$ by applying a gate $U$ to a target qubit $p$ conditioned on the state of $m$ control qubits.   These gates represent a basic logical building block  in the design of complex quantum algorithms, see {\sl e.g.} Refs.\cite{barenco1995,lanyon2009simplifying,qin2023error,zindorf2025}.
Here we encode the qubits in single-photon states on two coupled $M$-pods and we engineer driving protocols to control holonomically the evolution of the qubit states.
As specific examples we propose the realization of  the Toffoli gate~\cite{aharonov2003simple} and the quantum OR gate on a structure formed by two coupled $5$-pods which we call pentapods. Eventually we show  that similar structures can be used to implement the Deutsch quantum query algorithm~\cite{deutschquantum,portugal2024basicquantumalgorithms}.

\section{Photonic lattice and qubit encoding} 

We consider a chiral symmetric photonic lattice of eight waveguides coupled through evanescent hoppings $J_{i_p}$ with $i =b,c,d,x$, and $p=0,1$ shown in Fig.~\ref{fig:1}(b).
The lattice consists of two tripods, colored in blue and green, connected via the hoppings $J_{x_0}$ and $J_{x_1}$.  

The dynamics of photon propagation is  described by a Schr\"odinger equation $i\partial_z|\psi(z)\ra = H(z) |\psi(z)\ra $ with Hamiltonian 
\be H =  H_{0}  + H_{1} + H_{X}\ee where  the $0$,$1$ tripod Hamiltonian are given by 
$H_p = (J_{b_p} b_p^\dagger u_p + J_{c_p} c_p^\dagger u_p +  J_{d_p} d_p^\dagger u_p + \text{H.c.})$ with $p=0,1$,  while $H_{X}$ denotes the coupling between the two tripods $H_X = \big(J_{x_0} b_1^\dagger u_0 + J_{x_1} b_0^\dagger u_1 + \text{H.c.}\big)$.
Chirality ensures that, for any choice of the hopping parameters, the spectrum of $H$ consists of four degenerate modes 
$\mathcal{B} = \{|\psi_1\ra,\dots,|\psi_4\ra\}$ at energy $E_0=0$ and four non-degenerate non-zero modes. Chirality furthermore enforces that the zero modes lie only on the majority sublattice $\mathcal{M}=\{b_p,c_p,d_p\}_{p=0,1}$ -- {\it i.e.} they are zero on the minority sublattice $m= \{u_p\}_{p=0,1}$~\cite{ramachandran2017chiral}.

We focus on holonomic gates obtained by periodic and adiabatic  modulation of the couplings along $z$ {\sl i.e. } the propagation distance along the waveguides, which plays the role of time in the evolution process. At  $z=0$ only $J_{b_0}$ and $J_{b_1}$ are different from zero while all other couplings vanish. %$=J_x=J_{m_p}=0$.
In this initial configuration, the four orthonormal zero modes shown in Fig.~\ref{fig:1}(c), are 
\begin{equation}
    \begin{array}{cc}
    |\psi_{1}(0)\ra= \frac{|c_0\ra-|d_0\ra}{\sqrt{2}} \qquad 
    |\psi_{2}(0)\ra= \frac{|c_1\ra-|d_1\ra}{\sqrt{2}}  \\[0.2cm]
    |\psi_{3}(0)\ra= \frac{|c_0\ra+|d_0\ra}{\sqrt{2}} \qquad 
    |\psi_{4}(0)\ra=\frac{|c_1\ra+|d_1\ra}{\sqrt{2}}  \\
    \end{array}
    \label{eq:0_modes}
\end{equation}
yielding a natural encoding for two qubits in a single photon excitation.
We can therefore recast the logic states in Eq.~\eqref{eq:0_modes} as
\begin{equation} 
    \frac{|c_p\ra + (2s-1)  |d_p\ra}{\sqrt{2}}\equiv|s,p\ra
\label{eq:logic_states_t0}
\end{equation}
with  $p=0,1$ denoting the position and $s = 0,1$ the symmetry.
For the realization of controlled gates the position $p$ is the {\it target} qubit, while $s$ is the {\it control} qubit.

\section{Single and two-qubit gates}

We focus on the implementation of geometric gates by initializing the system in the degenerate subspace of zero modes  
$|\psi(0)\rangle = \sum_{\nu} c_{\nu} |\psi_\nu(0)\ra$  defined in Eq.~\eqref{eq:0_modes} and by manipulating adiabatically the coupling of the Hamiltonian $H$ along the $z$ coordinate following a cycle $\gamma$. 
Over the cycle $\gamma$ of duration $\lambda$,  the input state $|\psi(0)\rangle$ transforms as 
\be|\psi(\lambda)\rangle=W_\gamma(\lambda,0)|\psi(0)\rangle, \label{eq:ev}\ee %
where $W_\gamma(\lambda,0) =  P \exp\left[ i\oint_\gamma \Gamma_{z} dz \right]$ indicates the $U(4)$ holonomy,
$P$ denotes the path-ordering and  $[\Gamma_{z}]_{\nu\nu'} = \la \psi_{\nu}  | i\partial_z |\psi_{\nu'}\ra $ is the Wilczek-Zee connection.

As first highlighted by Zanardi and Rasetti \cite{ZANARDI199994},  non-Abelian holonomies can be exploited to implement a universal set gates. In particular, within the setup of Fig.~\ref{fig:1}(b), they allow for the implementation of rotations  and  phase gates acting on the symmetry quantum number $s$, respectively given by $V_\xi = e^{i \xi\sigma_y}\otimes \sigma_0$  and 
%$V_\varphi = \frac{1}{2}\left[\sigma_0+\sigma_z + e^{i\varphi}(\sigma_0-\sigma_z)\right]$
$V_\varphi = e^{i\frac{\varphi}{2} (\sigma_0-\sigma_z)}\otimes \sigma_0$. 

Both transformations can be generated holonomically through adiabatic manipulations acting on both tripods simultaneously and setting $H_T = 0$ following the route presented Refs~\cite{brosco2021nonabelian,pinske2020highly,pinske2022symmetry} for single tripod systems. 
 
For example modulating each tripod with the driving cycles shown in Fig.~\ref{fig:3}, one can realize the Hadamard gate and the phase gate, respectively -- see Appendix~\ref{app:1Qg} for more details. See Fig. \ref{fig:3}(a,b) for the corresponding final state tomography, extracted from the numerical data as $R_{s,s'}^{p,p'} = |\la s',p'|\psi_{s,p}(\lambda)\rangle|^2$.

\begin{figure}[t!]
	\centering
	\includegraphics[width=\columnwidth]{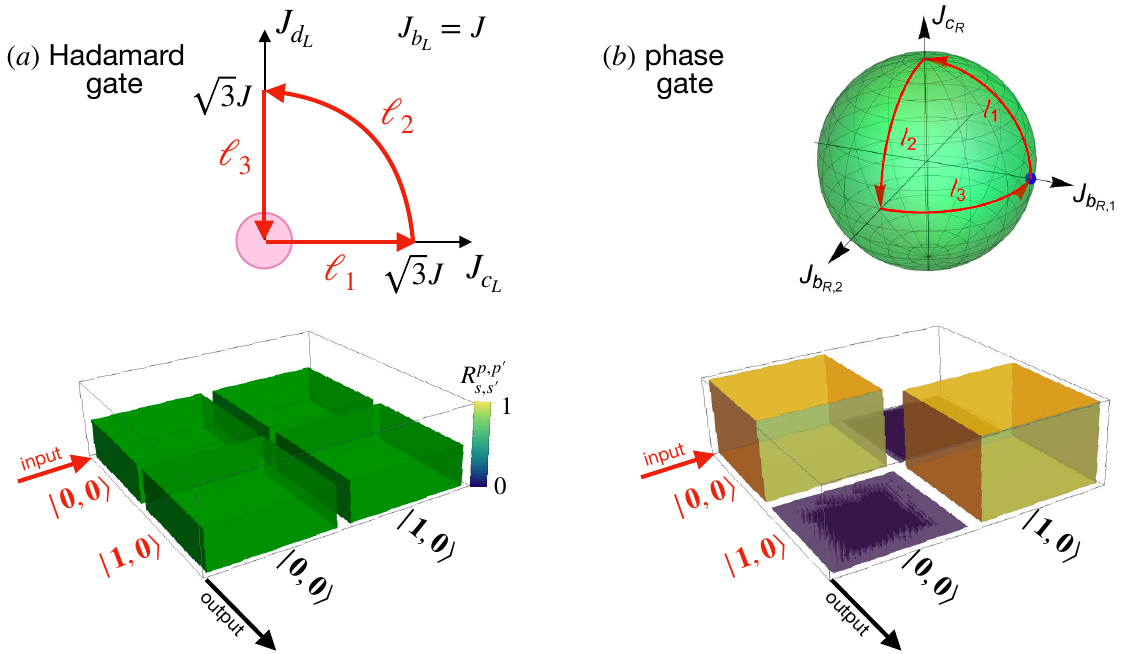}  
	\caption{
    (a) Driving cycle in the parameter space (the circle indicates the initial point) and quantum state tomography  for the single qubit Hadamard gate. 
	(b) Same as (a) for the single qubit phase gate.
	}
	\label{fig:3} 
\end{figure}

\begin{figure}[t!]
	\centering
    \includegraphics[width=\columnwidth]{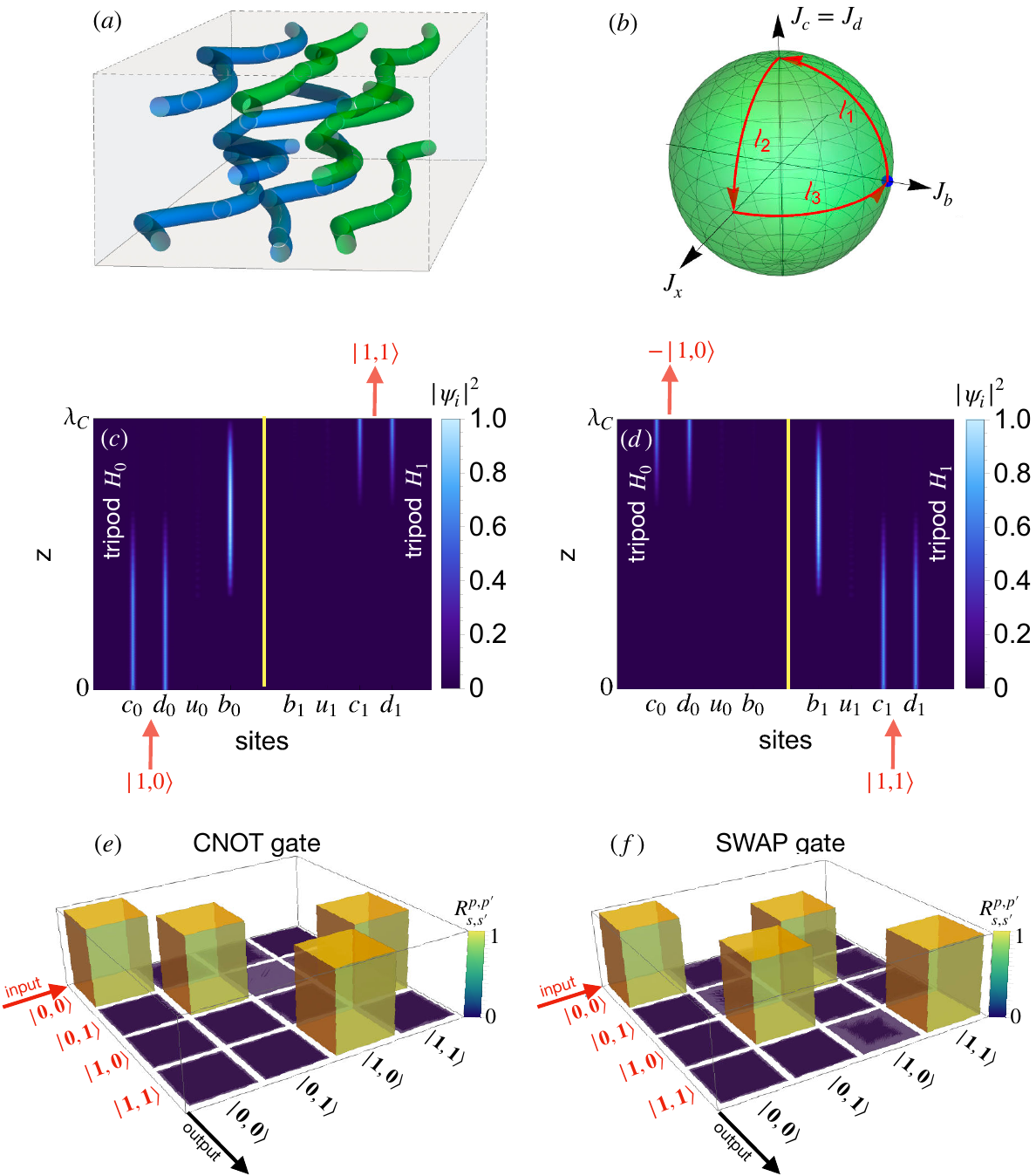}  
	\caption{
		(a) Illustration of the lattice pumped to implement a ${\rm CNOT}$ gate. The left tripod $p=0$ is colored in green while the right tripod $p=1$ is colored in blue. 
        (b) Driving cycle in the parameter space. The circle indicates the initial point. 
        (c,d) Propagation of $|1,0\ra$ and $|1,1\ra$ states respectively over one cycle period. The yellow lines separate the 0 and the 1 tripods.
        (e,f) Quantum state tomography for the ${\rm CNOT}$ gate and the ${\rm SWAP}$ gate respectively.   
}
	\label{fig:2} 
\end{figure}

For the four states system considered in this work,  the holonomic evolution along  suitably defined cycles also offers a natural playground to design two-qubit gates.
We start by focusing on the generation of conditional gates of the form:
\be 
C_\eta=|0\ra \la 0|\otimes \mathbb{I} + |1\ra \la1| \otimes U_\eta
\label{eq:gate}
\ee 
where $U_\eta = e^{i \eta\sigma_y}$
%$U_\eta = \cos\eta\, \sigma_0 + i \sin\eta\, \sigma_y$ 
is a rotation around the $y-$axis of the Bloch sphere and the angle $\eta$ depends only the geometry of the cycle and the structure of the Hilbert space -- see Appendix~\ref{app:2Qg} for more details.
An example is illustrated in Fig.~\ref{fig:2} where we demonstrate a ${\rm CNOT}$ gate -- {\it i.e.} a gate that swaps the target qubit $p$ if and only if the control qubit $s$ is $|1\ra$. The latter corresponds to setting $\eta=\pi/2$ -- hence, $U_\eta=i\sigma_y$ in Eq.\eqref{eq:gate}.

In our photonic array, this gate is achieved holonomically by setting $J_{x_0}=J_{x_1}\equiv J_{x}$, $J_{b_0}=J_{b_1}\equiv J_{b}$ and $J_{c_p}=J_{d_p}\equiv J_{c}$ for both $p=0,1$.  
The symmetry relations for $J_{c_p}=J_{d_p}$ imply that $|0,0\ra$ and $|0,1\ra$ are exact eigenstates of the array along the whole cycle $\gamma$  while the states $|1,0\ra$ and $|1,1\ra$ evolve adiabatically according to Eq.\eqref{eq:ev} leading $W_\gamma(\lambda,0)=C_{\pi/2}$ -- see Appendix~\ref{app:2Qg} for more details.

In Fig.~\ref{fig:2}(c,d) we show the evolution of the states $|1,0\ra$ and $|1,1\ra$ obtained by numerically solving the Schr\"odinger equation $i\partial_z|\psi(z)\ra = H(z) |\psi(z)\ra $. As predicted, the transformation flips the target qubit $p$ while preserving their control qubit $s=1$. In Fig.~\ref{fig:2}(e) we show the numerical quantum state tomography matrix $R$.  

\begin{figure*}[t!]
\centering\includegraphics[width=\textwidth]{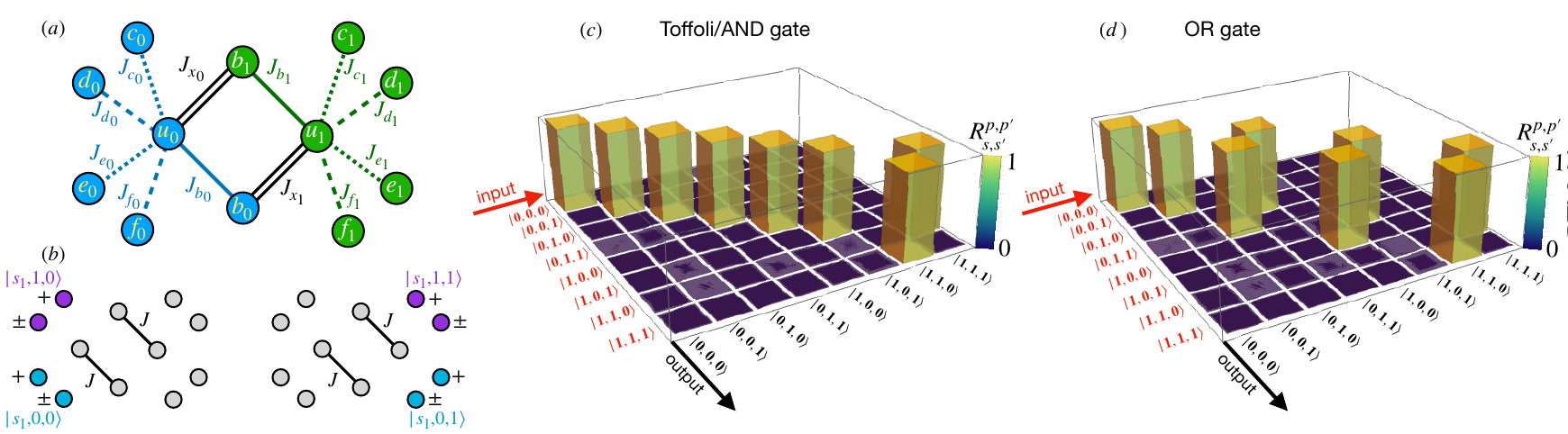}  
\caption{(a) Schematic of the lattice composed of left (0) and right (1) pentapods respectively colored in blue and green. The two tripods are connected via the hopping $J_{x_L}$ and $J_{x_R}$. 
(b) Three qubits encoding states $|b(s_1,s_2,p)\ra$ with two control qubits $s_1,s_2=\pm$ and a target $p=0,1$ qubits. At the initial parameters $J_{\ell_p}  =0$ for $\ell = c,d,e,f,x$ and $J_{b_p}=J$ for $p=L,R$ of the driving cycle, the states correspond to the eight zero-energy degenerate eigenstates.
(c) Quantum state tomography for the Toffoli gate. The driving cycle is shown in the top-right corner. 
(d) Same as (c) for a quantum version of the OR gate. }
\label{fig:4} 
\end{figure*} 

On the other hand, by setting $J_{c_1}=-J_{d_1}$ while keeping $J_{c_0}=J_{d_0}$ in the cycle in Fig.~\ref{fig:2}(b), the ${\rm CNOT}$ gate can be turned into a ${\rm SWAP}$ gate -- see Appendix~\ref{app:2Qg} for more details. This gate indeed, swaps the two qubits $|0,1\ra$ and $|1,0\ra$ while keeping $|0,0\ra$ and $|1,1\ra$ unchanged.
The numerically computed quantum state tomography is shown in Fig.~\ref{fig:2}(f). This implementation of the ${\rm SWAP}$ gate requires changing the sign of the hopping $J_{d_1}$ from positive to negative, which can be done experimentally by {\it e.g.} introducing an additional auxiliary waveguide between $u_1$ and $d_1$~\cite{keil2016universal,kremer2020square}, through a resistor applied to array~\cite{chang2021symmetry}, or by replacing certain waveguide with multi-orbital ones~\cite{roman2025observation,vicencio2025multiorbital}.

\section{Multi-controlled gates }

The array shown in Fig.~\ref{fig:1}(b) can be modified to host multi-controlled gates by replacing the two tripods % $H_0$ and $H_1$ 
with two $M$-pods. An $M$-pod is a system of $M+1$ waveguides with a star topology, where $M$ ones are connected to a central guide $u$, and it has $M-1$ zero-energy states~\cite{morris1983,pinske2020highly,pinske2022symmetry}. 
The Hamiltonians of the p-th $M$-pod  is 
\be H^{(M)}_p = \sum_{\ell \in [1,M]} J_{\ell,p} (a_{\ell,p}^\dagger u_{p} + \text{H.c.}) \quad p=0,1
\ee
with  $J_{\ell,p} $ denoting the coupling strengths of the $\ell$-th peripheral waveguide $a_{\ell,p}$ to the central central one $u_p$. %in the $p$-th $M$-pod . 
For $M=2^m +1$, % with $q=2^{m-1}$,
each $M$-pod has $2^m$ zero-energy states which can be used to encode $m$ qubit.\\ 
\noindent The array with two coupled $M$-pods has the following Hamiltonian
\be 
H^{(M)} = H^{(M)}_0 + H^{(M)}_1 + H_X
\ee
where the coupling Hamiltonian between the $M$-pods is $H_{X} = (J_{x_0} a_{1,1}^\dagger u_0 + J_{x_1} a_{1,0}^\dagger u_1 + \text{H.c.})$. The $2\cdot 2^{m}$ zero-energy states of the Hamiltonian $H^{(M)}$ yield the encoding of $m$ control qubits and one target qubit~\footnote{For comparison: (i) each tripod used in Fig.~\ref{fig:1}(b) has $M=3 = 2^{1} +1$ which result in a single ($m=1$) control qubit; while
(ii) each pentapod used in Fig.~\ref{fig:4}(a) has $M=5 =  2^{2} +1$ which result in two ($m=2$) control qubits.}.

Let us unfold this extension on the {\it Toffoli} gate. 
This gate is a doubly-controlled NOT -- {\it i.e.} it applies a NOT gate conditioned on $m=2$ control qubits, namely ${\rm CCNOT}=|0\ra \la0|\otimes \mathbb{I}_4 + |1\ra \la1| \otimes {\rm CNOT}$, which is an important component in quantum search algorithms~\cite{figgatt2017complete,Gidney2021howtofactorbit,chu2023scalable}. 
We realize the Toffoli gate by replacing each tripod in Fig.~\ref{fig:1}(b) with $5$-pods, as shown in Fig.~\ref{fig:4}(a). 
The Hamiltonian of each $5$-pods $H_p^{(5)}$, which we dub {\it pentapods}, is $H_p = (J_{b_p} b_p^\dagger +   J_{c_p} c_p^\dagger +  J_{d_p} d_p^\dagger +   J_{e_p} e_p^\dagger +  J_{f_p} f_p^\dagger) u_p + \text{H.c.}$.
The spectrum of the resulting array in Fig.~\ref{fig:4}(a) consists of eight degenerate modes $\{|\psi_j\rangle\}_{j=1}^8$ at energy $E_0=0$ and four non-degenerate modes, see appendix~\ref{app:lattice} and Ref~\cite{morris1983} for more details.

For $J_{b_0} =J_{b_1}= J$ and all other couplings vanishing, the orthonormal zero modes  are supported exclusively on to the disconnected external sites  of the two pentapods, namely, they are  linear combinations of the states $|c_0\ra$, $|d_0\ra$, $|e_0\ra$ and $|f_0\ra$ and $|c_1\ra$, $|d_1\ra$, $|e_1\ra$ and $|f_1\ra$. Choosing this initial configuration, limiting ourselves to single-excitation space, we can encode three qubits as follows \begin{equation}
\small
\begin{split}
   |s_1,s_2,p\ra&= \frac{s_2}{\sqrt{2}} (|c_p\ra+(2s_1-1)  |d_p\ra)   
  \\ &
   +   \frac{1-s_2}{\sqrt{2}}(|e_p\ra+(2s_1 -1) |f_p\ra)
   \end{split}
\label{eq:logic_states_t0_3q}
\end{equation}
where the {\it control} qubits are identified by the quantum numbers $s_1$ and $s_2$ while the {\it target} qubit is identified by  $p$.
The waveguides $b_0$ and $b_1$ are instead used only to couple the two pentapods.

By modulating the hopping terms $J_{b_p}$, $J_{c_p}$, and $J_{x_p}$ according to the driving cycle shown in Fig.~\ref{fig:2}(b), while setting all other couplings to zero ($J_{\ell_p}=0$ for $\ell=e,f$ and $p=0,1$), a Toffoli gate is implemented. The corresponding quantum state tomography $R$ for this gate is shown in Fig.~\ref{fig:4}(c). This gate is particularly notable because it realizes a reversible AND operation~\cite{nielsen2010quantum}. During the implementation of the Toffoli gate, the waveguides $e$ and $f$ remain decoupled from the rest of the circuit; however, they are activated to implement more complex logical operations.

As an example, Fig.~\ref{fig:4}(d) shows the quantum state tomography of a reversible OR gate~\cite{nielsen2010quantum}. This gate is obtained by composing three driving cycles of the form shown in Fig.~\ref{fig:2}(b), each time setting a different group of hopping amplitudes to zero. Specifically, the OR gate is implemented by applying the cycle in Fig.~\ref{fig:2}(b) three times with the following configurations: \\ 
(i) modulating $J_{b_p}$, $J_{c_p}$, and $J_{x_p}$ for $p=0,1$, while zeroing all other couplings; \\
(ii) modulating $J_{b_0}$, $J_{c_0}$, and $J_{x_0}$ on the $p=0$ tripod and $J_{e_1}$, $J_{f_1}$, and $J_{x_1}$ on the $p=1$ tripod, while zeroing the others; \\
(iii) modulating $J_{e_0}$, $J_{f_0}$, and $J_{x_0}$ on the $p=0$ tripod and $J_{b_1}$, $J_{c_1}$, and $J_{x_1}$ on the $p=1$ tripod, while zeroing the others.\\
These two gates shown in Fig.~\ref{fig:4}(c,d) are conditional gates of the form
\be 
C=\sum_{s_1,s_2} |s_1,s_2\ra \la s_1,s_2|\otimes V^{s_1,s_2} 
\label{eq:gate_2MC}
\ee 
% %
% \be
% e^{i q}
% \begin{pmatrix}
%     \cos\eta  & \sin\eta \\
%     -e^{i p} \sin\eta & e^{i p} \cos\eta
% \end{pmatrix}
% \quad |z|^2+|w|^2=1
% \ee
where $V^{s_1,s_2}$ is a $2\times 2$ unitary transformation which depend on $s_1$ and $s_2$ --   % and $V_\eta^{0,0} = \mathbb{I}$ -- 
hence, extending the conditional gate $C_\eta$ in Eq.~\eqref{eq:gate} from a single to a doubly controlled gate. 
The unitaries $V^{s_1,s_2}$ can be implemented by %activating the pentapods's waveguides by 
either composing driving  cycles that selectively involve a few waveguides only, as in our examples above, or driving cycles that involve all guides simultaneously.

\section{Implementation of a query problem}

We now show that the array in Fig.~\ref{fig:1}(b), together with the driving cycle in Fig.~\ref{fig:3} and \ref{fig:2}, can be used to implement quantum computing gate-models. A quantum gate-model consists of (i) a set of qubits, (ii) a sequence of quantum gates acting on them, and (iii) measurements performed on selected qubits.

A specific class of gate models is that of query models, which involve an oracle $U_f$ that encodes an unknown black-box function $f$ to be investigated.  Deutsch’s algorithm is a paradigmatic example of this class~\cite{portugal2024basicquantumalgorithms}. It determines whether a single-bit Boolean function $f:\{0,1\}\rightarrow\{0,1\}$ is constant, $f(0)=f(1)$, or balanced, $f(0)\neq f(1)$, using a single oracle query, whereas any classical strategy requires two. The four possible Boolean functions are summarized in Table~\ref{tab:1}, with $f_1$ and $f_4$ constant and $f_2$ and $f_3$ balanced.

 \begin{table}[h]
 \centering
% \begin{tabular}{ |P{0.5cm}|P{1cm}|P{1cm}|P{1cm}| P{1cm}|  }
  \begin{tabular}{ |c|c| c|c| c|  }
  \hline
 $s$ & $f_1(s) $ &$f_2(s)$& $f_3(s)$&$f_4(s)$\\
  \hline
 $0$ &$ 0$ &$0$&$1$&$1$\\
 $1$& $0$ &$1$&$0$&$1$\\
  \hline
 \end{tabular} 
 \caption{Table of the four Boolean functions.}
 \label{tab:1}
 \end{table}

\begin{figure}[t!]
	\centering
 \includegraphics[width=\columnwidth]{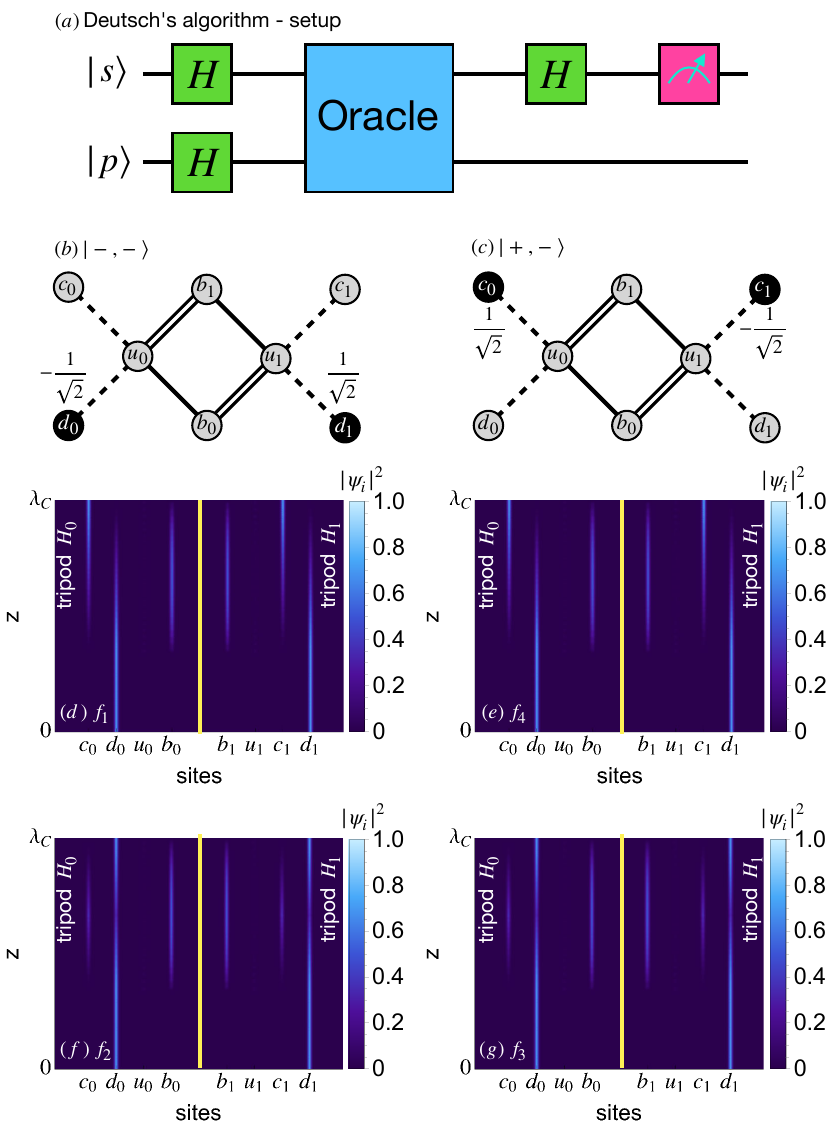}  
	\caption{ (a) Illustration of the Deutsch's algorithm. 
        (b,c) Superposition states $|+,-\ra$ and $|-,-\ra$. 
        (d,e) Propagation of $|-,-\ra$ over one cycle period for $\alpha_0=\pm 1$ and $\alpha_1=\pm 1$ which respectively implement the constant functions $f_1$ and $f_4$. 
        (f,g) Same as (d,e) for $\alpha_0=\pm 1$ and $\alpha_1=\mp 1$ which respectively implement the balanced functions $f_2$ and $f_3$. The yellow lines separate the 0 and the 1 tripods.
}
	\label{fig:5} 
\end{figure}

%\noindent
As presented in Ref.~\cite{portugal2024basicquantumalgorithms}, 
the oracle $U_f$ which implement the four Boolean functions is an operator acting on the two qubit states $|s,p\ra$ defined as $U_f(|s,p\ra) = |s,p\oplus f(s)\ra$where $\oplus$ is the XOR operation ({\it i.e.} the sum Modulo 2).
The query of the Deutsch's algorithm, as schematically recapped in Fig.~\ref{fig:5}(a), consists in considering a superposition state $|+,-\ra$ obtained by applying the Hadamard gates on both control and target qubits on an initial input $|0,1\ra$, as explicitly detailed in Appendix~\ref{app:DeuAlg}.  
%and  -- as shown in Fig.~\ref{fig:4}(a). 
The oracle $U_f$ %with a single query upon the state $|+,-\ra$  
discerns the functions in the two classes as 
%since $ U_f(|+,-\ra) = \pm |+,-\ra $ as 
\begin{equation}
  U_f(|+,-\ra)
=\left\{\begin{array}{@{}l@{}}
    (-1)^{f(0)}|+,-\ra \qquad 
    \text{constant} \\
%    f(0)\otimes f(1)=0 \\
      (-1)^{f(0)}|-,-\ra \qquad 
     \text{balanced} 
      %f(0)\otimes f(1)=1 
  \end{array}\right.\,
  \label{eq:query}
\end{equation}
The four operators $U_f$ can be obtained via four different combinations of the ${\rm CNOT}$ gate and single qubit NOT gate~\cite{portugal2024basicquantumalgorithms}. In our photonic array Fig.~\ref{fig:1}(b), these four oracle operators are implemented through proper combination of the driving cycles in Fig.~\ref{fig:3}(a,b) and Fig.~\ref{fig:2}(b). 
However, an alternative and less cumbersome way to implement the oracles comes from exploiting the symmetries of the array Fig.~\ref{fig:1}(b). 
Let us impose $J_{d_p} = \alpha_p J_{c_p}$ for $\alpha_p=\pm 1$ in each tripod that -- {\it i.e.}  $H_p = (J_{b_p} b_p^\dagger + J_{c_p} c_p^\dagger + \alpha_p J_{c_p} d_p^\dagger) u_p + \text{H.c.}$ 
for $p=0,1$. 
%The Hamiltonian $H = H_0+H_1+H_T$ is now dependent on $\sigma_0$ and $\sigma_1$. 
Consequently, the driving cycle in Fig.~\ref{fig:2}(b) yields four holonomies $W_{\alpha_0,\alpha_1}$ -- among which the holonomy corresponding to the ${\rm CNOT}$ can be obtained for $\alpha_0 = 1$ and $\alpha_1 = 1$ and the holonomy corresponding to the ${\rm SWAP}$ can be obtained for $\alpha_0 = 1$ and $ \alpha_1 = -1$. 
In our photonic array in Fig.~\ref{fig:1}(b), the superposition states read $|-,-\ra = \frac{1}{\sqrt{2}}(-|d_0\ra + |d_1\ra)$ and $|+,-\ra = \frac{1}{\sqrt{2}}(|c_0\ra - |c_1\ra)$ -- respectively shown in Fig.~\ref{fig:5}(b,c). 
In Fig.~\ref{fig:5}(d-g) we show the single queries which distinguish the identity functions $f_1,f_4$ (shown in Fig.~\ref{fig:5}(d,e)) from the balanced functions $f_2,f_3$ (shown in Fig.~\ref{fig:5}(f,g)).

Let us conclude by conjecturing that the proposed extension of the array from interconnected tripods to interconnected $M$-pods may pave the way to generalize this procedure. 
Indeed, setting $M=2^m+1$ could allow the photonic implementation of the Deutsch-Jozsa quantum algorithm via non-Abelian holonomies. 
This is an algorithm that efficiently distinguish $m$-bits binary functions  $f(s): \{0,1\}^m\rightarrow \{0,1\}$ between constant (the function returns the same value for all $2^m$ inputs) and balanced (the function returns 0 for half $2^{m-1}$ inputs and 1 for the remaining half).

\section{Conclusions}\label{sec:conclusion}
Recent work on non-Abelian Thouless pumping \cite{brosco2021nonabelian,danieli2025nonabelian,Sun2022}, non-Abelian braiding \cite{yang2024,sun2025,Chen2025,chen2019non}, and quantum walks \cite{danieli2024parity} has demonstrated that holonomies in photonic waveguide arrays can be used to control light propagation, enabling the engineering of topological effects.

\noindent In the present work, we explore the possibility of encoding the state of single and multiple qubits in the position of a single photon within a waveguide array, and we investigate how non-Abelian holonomies can be used to implement different quantum gates  realizing an holonomic version of linear optic quantum computing protocols \cite{kok2017}.\\
\noindent Our control strategy is based on the paradigm of \textit{adiabatic population transfer}, similar to that employed in STIRAP atomic protocols. Accordingly, we focus on lattices with the structure of single or coupled M-pods, or star graphs, where a single site is coupled to $M$ peripheral sites. Within these structures, we show that it is possible to engineer manipulations in which the structure of the single-photon wavefunction on a single $M$-pod  with $M=2^{m}+1$ acts as $m$ control qubits, while the placement on one or another M-pods serves as the target.
Within this conceptual framework, we discuss the implementation of
(i) Single-qubit gates, such as the Hadamard and phase gates, and
(ii) Multi-qubit gates, including the CNOT and other multi-controlled gates.

We further show that the proposed photonic array can realize notable quantum query algorithms, such as the Deutsch algorithm, where the adiabatic evolution operator associated with a holonomy acts as the oracle.
This holonomic implementation of quantum query problems reframes both the oracle and queries in terms of geometric holonomies. Our approach thus opens a novel direction for leveraging non-Abelian holonomies in quantum algorithms and quantum computing.

\section*{Acknowledgement}

This work was co-funded by Project PNRR MUR
PE 0000023-NQSTI, Project PNRR MUR project CN
00000013-ICSC, Fondazione Cariplo grant 2023-2594, and the European Union (HORIZON-ERC-2023-ADG HYPERSPIM project Grant No. 101139828). Views and opinions expressed are however those of the author(s) only and do not necessarily reflect those of the European Union of the European Research Council. Neither the European Union nor the granting authority can be held responsible for them.

\section*{Data availability} %ATA AVAILABILITY
The data that support the findings of this article are not
publicly available. The data are available from the authors
upon reasonable request. 

% \bibliography{Qgates}

% \clearpage

\appendix
%\section{Supplementary material}

\section{Eight-sites array and holonomies}\label{app:lattice}

The Hamiltonian of the photonic array written explicitly reads 
\be
\begin{split} 
H &=  H_{0}  + H_{1} + H_{T}  \\
&= (J_{b_0} b_0^\dagger +   J_{c_0} c_0^\dagger +  J_{d_0} d_0^\dagger) u_0 \\
&+ (J_{b_1} b_1^\dagger +   J_{c_1} c_1^\dagger +  J_{d_1} d_1^\dagger) u_1 \\
&+ J_{x_0} b_1^\dagger u_0 + J_{x_1} b_0^\dagger u_1 + \text{H.c.}
\end{split}
\label{eq:H_app}
\ee
which has matrix 
\be
{\small H=\begin{pmatrix}
	0 & J_{b_0} & J_{c_0} & J_{d_0} & 0 & J_{x_0} & 0 & 0 \\
	J_{b_0} & 0 & 0 & 0 & J_{x_1} & 0 & 0 & 0 \\
	J_{c_0} & 0 & 0 & 0 & 0 & 0 & 0 & 0 \\
	J_{d_0} & 0 & 0 & 0 & 0 & 0 & 0 & 0   \\
	0 & J_{x_1} & 0 & 0 & 0 & J_{b_1} & J_{c_1} & J_{d_1} \\
	J_{x_0} & 0 & 0 & 0 & J_{b_1} & 0 & 0 & 0 \\
	0 & 0 & 0 & 0 & J_{c_1} & 0 & 0 & 0 \\
	0 & 0 & 0 & 0 & J_{d_1} & 0 & 0 & 0
\end{pmatrix} }
\label{eq:8sites}
\ee
Due to chiral symmetry, this array has four degenerate eigenenergies at $E_0 = 0$ for any value of the parameters. The other four non-degenerate eigenenergies are 
\be
\footnotesize
\begin{split}
&\quad E=\\
&\pm \frac{1}{\sqrt{2}}\Bigg\{J_{b_0}^2 + J_{b_1}^2 + J_{c_0}^2 + J_{c_1}^2 + J_{d_0}^2 + J_{d_1}^2 + J_{x_0}^2 + J_{x_1}^2 \\
    & \pm  \bigg[ (J_{b_0}^2 + J_{b_1}^2 + J_{c_0}^2 + J_{c_1}^2 + J_{d_0}^2 + J_{d_1}^2  + J_{x_0}^2 + J_{x_1}^2)^2 \\
    & +  4 \Big( 2 J_{b_0} J_{b_1} J_{x_0} J_{x_1} - (J_{c_0}^2 + J_{d_0}^2 + J_{x_0}^2) (J_{c_1}^2 + J_{d_1}^2 + J_{x_1}^2) \\
    & -  J_{b_1}^2 (J_{c_0}^2 + J_{d_0}^2 ) - J_{b_0}^2 (J_{b_1}^2 + J_{c_1}^2 + J_{d_1}^2)
    \Big)\bigg]^{\frac{1}{2}}  \Bigg\}^{\frac{1}{2}}   
\end{split}
\label{eq:energies_app}
\ee

\noindent
Chirality constricts the four eigenstates onto the majority subarray $\mathcal{M}=\{b_0,c_0,d_0,b_1,c_1,d_1 \}$. 
Degeneracy implies that there exist arbitrary choices of orthonormal basis $\mathcal{B} = \{|\psi_1\ra,\dots,|\psi_4\ra\}$ of the $E_0 = 0$ degenerate subspace.  
In the following computation of the Wilczek-Zee connection 
\be
[\Gamma_{z}]_{\nu\nu'} = \la \psi_{\nu}  | i\partial_z |\psi_{\nu'}\ra 
\label{eq:WZ_app}
\ee 
used to obtain  the holonomy 
\be 
W(z_0,z_1) = e^{ i\theta_{\rm d}} P \exp\left[ i\int_{z_0}^{z_1} \Gamma_{z} dz \right] 
\label{eq:holo_app} 
\ee
we specify basis $\mathcal{B}$ used to compute $\Gamma_{z}$ for each gate. This does not affect the calculation nor loose generality, but avoid us to report four cumbersome eigenstates.

\section{Single qubit gates}\label{app:1Qg}

\begin{figure}[htbp]
	\centering
	\includegraphics[width=0.375\columnwidth]{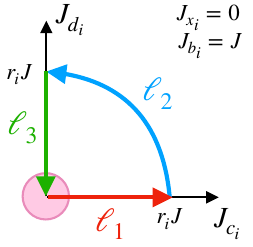}  
	\caption{Driving cycle $\gamma_{\sigma_y}$ split in three branches $\ell_1$ (red), $\ell_2$ (blue) and $\ell_3$ (green). }
	\label{fig:cycle_sigmaY_app}
\end{figure}

To implement single qubit Pauli gates represented by $\sigma_y$ ad $\sigma_z$, we set $J_{x_0}=J_{x_1}\equiv 0$ for all $z\in\mathbb{R}$.   
In this set-up, the four non-zero modes in Eq.~\eqref{eq:energies_app} reduce to $E=\left\{\pm \sqrt{J_{c_0}^2+J_{d_0}^2 + J_{b_0}^2},\pm \sqrt{J_{c_1}^2+J_{d_1}^2 + J_{b_1}^2}\right\}$. 
An orthonormal basis of zero-modes for this specific configuration is:
\begin{eqnarray}
  |\psi_{1}^\text{y}(z)\ra&=&  \frac{J_{d_0}|c_0\ra- J_{c_0}|d_0\ra}{\delta_0} \\ \nonumber
     |\psi_{2}^\text{y}(z)\ra&=&\frac{J_{d_1}|c_1\ra- J_{c_1}|d_1\ra}{\delta_1} \\ \nonumber
   |\psi_{3}^\text{y}(z)\ra&=&  \frac{J_{b_0}( J_{c_0}|c_0\ra+ J_{d_0}|d_0\ra) -  \delta_0^2 |b_0\ra }{\delta_0 \Delta_0}  \\ \nonumber
   |\psi_{4}^\text{y}(z)\ra&=& \frac{J_{b_1}( J_{c_1}|c_1\ra+ J_{d_1}|d_1\ra) -  \delta_1^2 |b_1\ra }{\delta_1 \Delta_1}\nonumber
\label{eq:sigmaY_states_app}
\end{eqnarray} 
where $\delta_p = \sqrt{J_{c_p}^2 + J_{d_p}^2}$ and $\Delta_p = \sqrt{J_{c_p}^2 + J_{d_p}^2+ J_{b_p}^2}$ for $p=0,1$. 
At the starting point of the evolution where the parameters are $J_{c_p}=J_{d_p}=0, J_{b_p}=J$, these eigenstates reduce to the encoding states $|s,p\ra$ for  $p=0,1$ and  $s = 0,1$ as in Eq.~\eqref{eq:cnot_logic_states_app}. %\\ \\
Let us compute the auxiliary terms $A_{\nu\nu'}$ useful to obtain the connection $\Gamma_z$ in Eq.~\eqref{eq:G_gen_app}, here written as 
\begin{equation}
\label{eq:Aaux_sigmaY_app}
\begin{split}
A_{\nu\nu'} &= i \Big\{  \langle \psi_{\nu} | \partial_{J_{c_0}} |\psi_{\nu'} \rangle, \langle \psi_{\nu}| \partial_{J_{d_0}}  | \psi_{\nu'} \rangle , \langle \psi_{\nu}|\partial_{J_{b_0}} | \psi_{\nu'} \rangle, \\
& \qquad \langle \psi_{\nu} | \partial_{J_{c_1}} |\psi_{\nu'} \rangle, \langle \psi_{\nu}| \partial_{J_{d_1}}  | \psi_{\nu'} \rangle , \langle \psi_{\nu}|\partial_{J_{b_1}} | \psi_{\nu'} \rangle \Big\} 
\end{split}
\end{equation}
The only relevant terms are $A_{13}$ and $A_{24}$, which read:
\begin{equation}
%\footnotesize
\begin{split}
A_{13} &=   \frac{i J_{b_0}}{\delta_0^2 \Delta_0} \left\{ J_{c_0},-J_{d_0},0,0,0,0  \right\}   \\ 
A_{24} &=   \frac{i J_{b_1}}{\delta_1^2 \Delta_1} \left\{ 0,0,0,J_{c_1},-J_{d_1},0  \right\}   
\end{split}
\label{eq:Gaux_sigmaY_app}
\end{equation}
with $A_{31}  = -A_{13}$ and $A_{42}  = -A_{24}$. The Wilczek-Zee connection then reads
\begin{equation}
\begin{split}
%\footnotesize
\Gamma_z 
&=  \frac{ J_{b_0} (  J_{d_0}  \dot{J}_{c_0} - J_{c_0} \dot{J}_{d_0})  }{\delta_0^2 \Delta_0}  \sigma_y  \otimes \frac{\sigma_0 +  \sigma_z}{2}  \\
& + \frac{ J_{b_1} (  J_{d_1}  \dot{J}_{c_1} - J_{c_1} \dot{J}_{d_1})  }{\delta_1^2 \Delta_1}  \sigma_y  \otimes \frac{\sigma_0 -  \sigma_z}{2}
\end{split}  
\label{eq:G_sigma_Y_app}
\end{equation}
For closed paths $\gamma_p$ in the parameter space, the path integrals yield $\xi_0=\oint_{\gamma_0} [\Gamma_z ]_{13} d\gamma$ and $\xi_1=\oint_{\gamma_1} [\Gamma_z ]_{24} d\gamma$. In Eq.~\eqref{eq:holo_app}, this results in:
\begin{equation}
\small 
\begin{split}
W(\gamma) &= \begin{pmatrix} 
   \cos\xi_0&0&\sin\xi_0&0 \\
   0&\cos\xi_1 &0&\sin\xi_1 \\
   -\sin\xi_0&0&\cos\xi_0&0 \\
   0&-\sin\xi_1&0&\cos\xi_1 
   \end{pmatrix} \\
&= \left[ \cos\xi_0 \sigma_0 + i \sin\xi_0 \sigma_y \right] \otimes \frac{\sigma_0 + \sigma_z}{2} \\
&+ \left[ \cos\xi_1 \sigma_0 + i \sin\xi_1 \sigma_y  \right] \otimes \frac{\sigma_0 - \sigma_z}{2}  
\end{split}
\label{eq:holo_sigmaY_angle}
\end{equation}
These are two independent $\sigma_y$ rotations for single qubits in each tripod. 
Loops of the kind shown in Fig.~\ref{fig:cycle_sigmaY_app}, split in three separate branches $\ell_1$, $\ell_2$ and $\ell_3$ respectively colored in red, blue and green, allow to implement, among the others, the Hadamard and the NOT gates.

\noindent
The first and third branches $\ell_1$ and $\ell_3$ yield zeros in Eq.~\eqref{eq:G_sigma_Y_app} as along them $J_{c_p}$ and $J_{d_p}$ are respectively zero. Along $\ell_2$, $J_{b_p}=J$ while $J_{c_p} = r_p J\cos\theta$ and $J_{d_p}=r_pJ\sin\theta$ for $\theta\in[0,\frac{\pi}{2}]$. The prefactors of the Pauli matrix tensor products in Eq.~\eqref{eq:G_sigma_Y_app} then reduce to $\frac{ 1  }{\sqrt{r_p^2+1}}$. Hence, in Eq.~\eqref{eq:holo_sigmaY_angle} the angles are $\eta_p = \frac{ \pi }{2\sqrt{r_p^2+1}}$. 
The Hadamard gate is achieved by choosing $r_p=\sqrt{3}$, which implies $\eta_p = \frac{ \pi }{4}$. A NOT gate obtained for $\eta_p = \frac{ \pi }{2}$ in Eq.~\eqref{eq:holo_sigmaY_angle}  can be achieved by  either extending the quarter circle to a half circle ({\it i.e.} for $\theta\in[0, \pi]$) or by simply repeating the loop $\gamma_{\sigma_y}$ in Fig.~\ref{fig:cycle_sigmaY_app} twice. 
The holonomy in Eq.~\eqref{eq:holo_sigmaY_angle} for  $\xi_0=\frac{\pi}{4}$ and $\xi_1=\frac{\pi}{2}$ hence reduces to
\begin{equation}
\small 
\begin{split}
W &= \frac{1}{\sqrt{2}} \begin{pmatrix} 
   1&0&1&0 \\
   0& 0 &0& \sqrt{2} \\
   -1&0&1 &0 \\
   0&-\sqrt{2}&0& 0  
   \end{pmatrix} \\
&= \frac{ \sigma_0 + i \sigma_y}{\sqrt{2}} \otimes \frac{\sigma_0 + \sigma_z}{2} +  i  \sigma_y  \otimes \frac{\sigma_0 - \sigma_z}{2}  
\end{split}
\label{eq:holo_sigmaY_2}
\end{equation}

\begin{figure}[htbp]
	\centering
	\includegraphics[width=\columnwidth]{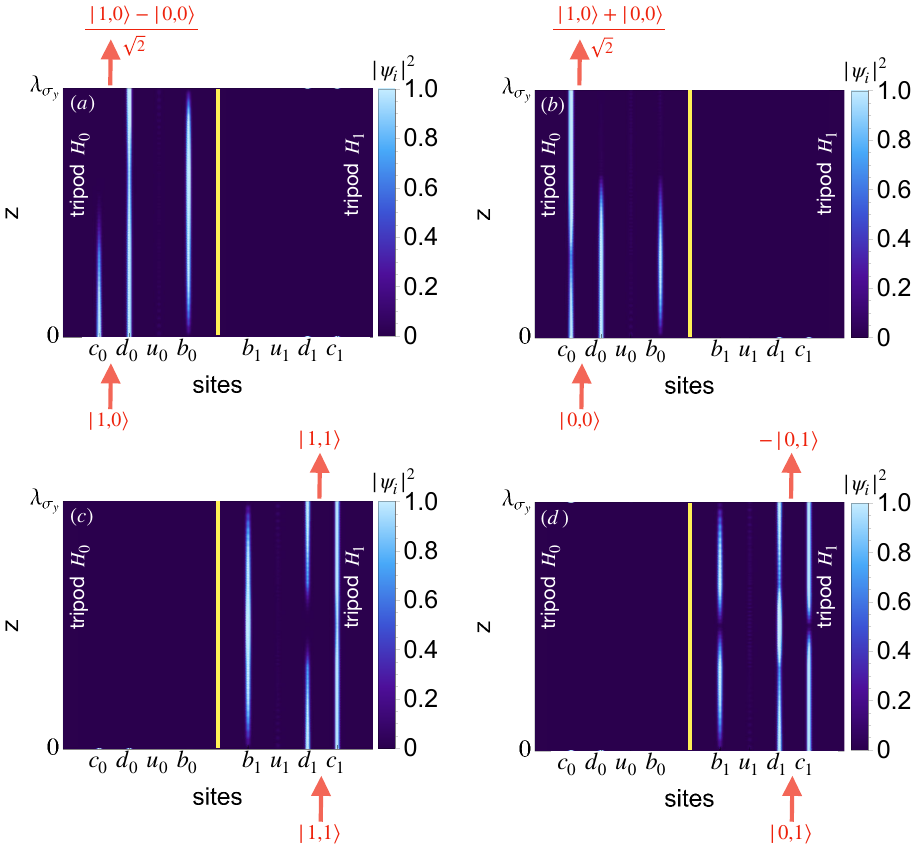} 
	\caption{(a,b) Evolution of the $|1,0\ra$ and $|0,0\ra$ respectively via cycle in Fig.~\ref{fig:cycle_sigmaY_app} with $\xi_0=\frac{\pi}{4}$ in the tripod $H_0$. 
    (c,d) Evolution of the $|1,1\ra$ and $|0,1\ra$ respectively via cycle in Fig.~\ref{fig:cycle_sigmaZ_app} in the tripod $H_1$. 
    The yellow lines separate the 0 and the 1 tripods.  }
	\label{fig:cycle_sigmaY_evo_app}
\end{figure}

In Fig.~\ref{fig:cycle_sigmaY_evo_app}, we show the evolution of the states $|1,0\ra$ [panel(a)] and $|0,0\ra$ [panel(b)] in the tripod $H_0$ via the cycle in Fig.~\ref{fig:cycle_sigmaY_app} defined for $\xi_0=\frac{\pi}{4}$. 
The two tripods are disjoint as the connecting Hamiltonian $H_T = 0$.  
Both panels show that for $\xi_0=\frac{\pi}{4}$, the initial states end respectively in a superposition  $\frac{|1,0\ra -|0,0\ra }{\sqrt{2}}$ and $\frac{|1,0\ra +|0,0\ra }{\sqrt{2}}$, which correspond to a $\frac{\pi}{4}$ rotation.

% In Fig.~\ref{fig:cycle_sigmaY_evo_app}, we show the evolution of the encoding states $|s,p\ra$ for $s=0,1$ and $p=0,1$ via the cycle in Fig.~\ref{fig:cycle_sigmaY_app}. The two tripods are disjoint as the connecting Hamiltonian $H_T = 0$. For the tripod (0) we set $\xi_0=\frac{\pi}{4}$ while in the tripod (1) we set $\xi_1=\frac{\pi}{2}$. 
% Both panels show that for $\xi_0=\frac{\pi}{4}$) the states $|s,0\ra$ end in a superposition  $\frac{|s,0\ra -(2s-1) |1-s,0\ra }{\sqrt{2}}$ while for $\xi_1=\frac{\pi}{2}$ the two states $|s,1\ra$ switches. 

%

\begin{figure}[htbp]
	\centering
	\includegraphics[width=0.45\columnwidth]{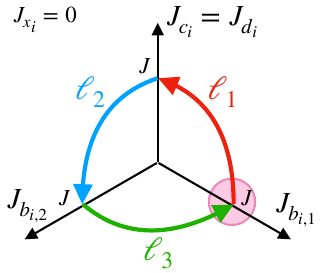}  
	\caption{Driving cycle $\gamma_{\sigma_z}$ split in three branches $\ell_1$ (red), $\ell_2$ (blue) and $\ell_3$ (green). }
	\label{fig:cycle_sigmaZ_app}
\end{figure}

%\noindent
To perform a phase qubit $\sigma_z$, we assume $J_{c_p} = J_{d_p}$ in both tripods $p=0,1$ and turn the hoppings $J_{b_p}$  complex, {\it i.e.}  $J_{b_p} = J_{b_{i,1}}+e^{i\phi_p} J_{b_{i,2}}$ for phases $\phi_p$. 
An orthonormal basis of zero-modes, for this specific configuration, reads:
\begin{eqnarray}
  |\psi_{1}^\text{z}(z)\ra&=&  \frac{|c_0\ra-|d_0\ra}{\sqrt{2}}\\ \nonumber
  |\psi_{2}^\text{z}(z)\ra&=&   \frac{|c_1\ra-|d_1\ra}{\sqrt{2}}  \\ \nonumber
   |\psi_{3}^\text{z}(z)\ra&=&\frac{J_{b_0}  (|c_0\ra+ |d_0\ra) -  J_{c_0} |b_1\ra }{\epsilon_0}  \\ \nonumber
   |\psi_{4}^\text{z}(z)\ra&=& \frac{J_{b_1}(|c_1\ra+ |d_1\ra) -  J_{c_1} |b_0\ra }{\epsilon_1}\nonumber
\label{eq:sigmaZ_states_app}
\end{eqnarray} 
with $\epsilon_p = \sqrt{2(2 J_{c_p}^2 + |J_{b_p}|^2)}$. 
At the initial point of the cycle where the parameters are $J_{c_p}=0, J_{b_{i,1}}=J $ and $J_{b_{i,2}}=0$, these eigenstates reduce to the encoding states $|s,p\ra$ as in Eq.~\eqref{eq:logic_states_t0}. 
 The auxiliary terms $A_{\nu\nu'}$ useful to obtain the connection $\Gamma_z$ in Eq.~\eqref{eq:G_gen_app}, here are: 
\begin{equation}
\label{eq:Aaux_sigmaY_app}
\begin{split}
A_{\nu\nu'} &= i \Big\{  \langle \psi_{\nu} | \partial_{J_{c_0}} |\psi_{\nu'} \rangle, \langle \psi_{\nu}| \partial_{J_{b_{0,1}}}  | \psi_{\nu'} \rangle , \langle \psi_{\nu}|\partial_{J_{b_{0,2}}} | \psi_{\nu'} \rangle, \\
& \qquad \langle \psi_{\nu} | \partial_{J_{c_1}} |\psi_{\nu'} \rangle, \langle \psi_{\nu}| \partial_{J_{b_{1,1}}}  | \psi_{\nu'} \rangle , \langle \psi_{\nu}|\partial_{J_{b_{1,2}}} | \psi_{\nu'} \rangle \Big\} 
\end{split}
\end{equation}
The only relevant terms are $A_{33}$ and $A_{44}$, which read
\begin{equation}
%\footnotesize
\begin{split}
A_{33} &=   \frac{ \sin \phi_0}{\epsilon_0^2}  \left\{ 0 , J_{b_{0,2}} ,-J_{b_{0,1}},0,0,0,0  \right\}   \\ 
A_{44} &=   \frac{ \sin \phi_1}{\epsilon_1^2} \left\{ 0,0,0,0, J_{b_{1,1}},-J_{b_{1,2}}  \right\}   
\end{split}
\label{eq:Gaux_sigmaZ_app}
\end{equation}
 The Wilczek-Zee connection then reads
\begin{equation}
%\footnotesize
\small
\begin{split}
\Gamma_z 
&=  \frac{  \sin \phi_0 ( J_{b_{0,2}} \dot{J}_{b_{0,1}} - J_{b_{0,1}} \dot{J}_{b_{0,2}}  )  }{\epsilon_0^2}  \frac{\sigma_0 -  \sigma_z}{2}   \otimes \frac{\sigma_0 +  \sigma_z}{2}  \\
& + \frac{ \sin \phi_1 ( J_{b_{1,2}} \dot{J}_{b_{1,1}} - J_{b_{1,1}} \dot{J}_{b_{1,2}}  )  }{\epsilon_1^2}  \frac{\sigma_0 -  \sigma_z}{2}   \otimes \frac{\sigma_0 -  \sigma_z}{2}
\end{split}  
\label{eq:G_sigma_Z_app}
\end{equation}
For a closed path $\gamma$ in the parameter space, the path integrals yield $\varphi_0=\oint_{\gamma_0} [\Gamma_z ]_{33} d\gamma$ and $\varphi_1=\oint_{\gamma_1} [\Gamma_z ]_{44} d\gamma$. In Eq.~\eqref{eq:holo_app}, this results 
\begin{equation}
\small
\begin{split}
W(z_0,z_1) & = 
   \begin{pmatrix} 
   1&0&0&0 \\
   0&1&0&0 \\
   0&0&e^{i\varphi_0} &0 \\
   0&0&0&e^{i\varphi_1}
   \end{pmatrix} \\
&= \frac{\sigma_0 - \sigma_z}{2} \otimes  \left[ e^{i\varphi_0}  \frac{\sigma_0 + \sigma_z}{2} 
+ e^{i\varphi_1} \frac{\sigma_0 - \sigma_z}{2} \right] \\
&+\frac{\sigma_0 + \sigma_z}{2} \otimes \sigma_0  
\end{split}
\label{eq:holo_sigmaZ_angle}
\end{equation}
A typical loop $\gamma_{\sigma_z}$ that provides a phase shift gate is shown in Fig.~\ref{fig:cycle_sigmaZ_app}, split in three branches $\ell_1$, $\ell_2$ and $\ell_3$ colored in red, blue and green, respectively.
\noindent
The first and the third branches $\ell_1$ and $\ell_3$ yield zeroes in Eq.~\eqref{eq:G_sigma_Z_app} as along them $J_{b_{i,1}}$ and $J_{b_{i,2}}$ are respectively zero. Along $\ell_2$ the hopping $J_{c_p} = 0 = J_{d_p}$, which implies that the prefactors of the Pauli matrix tensor products in Eq.~\eqref{eq:G_sigma_Z_app} are the derivative in $z$ of the complex parameters -- {\it i.e.} $\partial_z J_{b_L} = [\Gamma_x]_{33}$ and  $\partial_z J_{b_R} = [\Gamma_x]_{44}$. 
In this branch $\ell_2$, where $J_{b_{i,2}} = J\cos\theta$ and $J_{b_{i,1}}=J\sin\theta$ for $\theta\in[0,\frac{\pi}{2}]$, the path integrals reduces to $\varphi_0=\oint_\gamma [\Gamma_z ]_{33} d\gamma = \phi_L$ and $\varphi_1=\oint_\gamma [\Gamma_z ]_{44} d\gamma = \phi_R$. 
Hence, the holonomy in Eq.~\eqref{eq:holo_sigmaZ_angle} represent two relative phase rotation gates, one in each tripod $H_L$ and $H_R$. These rotations reduce to two $\sigma_z$ gates, one per tripod, by choosing $\phi_p = \pi$, whose matrix is
\begin{equation}
%\small
\begin{split}
W(z_0,z_1) & = 
   \begin{pmatrix} 
   1&0&0&0 \\
   0&1&0&0 \\
   0&0&-1 &0 \\
   0&0&0&-1
   \end{pmatrix} 
=  \sigma_z \otimes \sigma_0  
\end{split}
\label{eq:holo_sigmaZ_2}
\end{equation}  

In Fig.~\ref{fig:cycle_sigmaY_evo_app}, we show the evolution of the states $|1,1\ra$ [panel(c)] and $|0,1\ra$ [panel(d)] in the tripod $H_1$ via the cycle in Fig.~\ref{fig:cycle_sigmaZ_app}. 
The two tripods are disjoint as the connecting Hamiltonian $H_T = 0$.  
These panels show that for $\xi_0=\frac{\pi}{4}$, the initial state $|1,1\ra$ is mapped into itself, while $|0,1\ra$ flips sign.

\section{Two qubits gates}\label{app:2Qg}

Let us discuss the ${\rm CNOT}$ and the ${\rm SWAP}$ gate. To realize the former one, we consider a fully symmetric structure where $J_{x_0}=J_{x_1}\equiv J_{x}$, $J_{b_0}=J_{b_1}\equiv J_{b}$ and $J_{c_0}=J_{c_1}=J_{d_0}=J_{d_1}\equiv J_{c}$ for which the Hamiltonian reduces to:
\begin{equation}
\begin{split}
   H(z) &= J_{b}(b_0^\dagger u_0+b_1^\dagger u_1)+J_{x}(b_1^\dagger u_0+b_0^\dagger u_1) \\
   &+ J_{c}[( c_0^\dagger+d_0^\dagger)u_0+(c_1^\dagger+d_1^\dagger)u_1] + \text{H.c.}
   \end{split}
   \label{eq:Hcnot_app}
\end{equation}
In this set-up, the four non-zero modes in Eq.~\eqref{eq:energies_app} reduce to $E=\pm \sqrt{2J_c^2+(J_x \pm J_b)^2}$. 
We introduce an orthonormal basis of zero-modes for this specific driving cycle -- hence, we denote the basis with an apex c
\begin{eqnarray}
  |\psi_{1}^\text{c}(z)\ra&=&  \frac{|c_0\ra-|d_0\ra}{\sqrt{2}} \\ \nonumber
   |\psi_{2}^\text{c}(z)\ra&=&  \frac{|c_1\ra-|d_1\ra}{\sqrt{2}} \\ \nonumber
   |\psi_{3}^\text{c}(z)\ra&=&
    \frac{J_b ( 2J_c^2 -  J_x^2 +  J_b^2 )  |s_0\ra +
J_x (2J_c^2  +  J_x^2 -  J_b^2 )  |s_1\ra}{R_3} \\
 &+&    \frac{\sqrt{2} J_c (J_b^2 + 2 J_c^2 + J_x^2) |b_0\ra -  2 \sqrt{2} J_x J_b J_c  |b_1\ra }{R_3} \\ \nonumber 
    |\psi_{4}^\text{c}(z)\ra&=&
   \frac{ J_x |s_0\ra +  J_b |s_1\ra - \sqrt{2}J_c |b_1\ra }{R_4} \\ \nonumber
\label{eq:cnot_states_app}
\end{eqnarray} 
where $|s_0\ra= \frac{|c_0\ra+|d_0\ra}{\sqrt{2}}$, $|s_1\ra= \frac{|c_1\ra+|d_1\ra}{\sqrt{2}}$, 
$N_{\pm}= \sqrt{2J_c^2+(J_x\pm J_b)^2}$,  $R_4 = \sqrt{ 2J_c^2+ J_x^2 + J_b^2 }$ and $R_3 = R_4 N_+ N_- $. 
At the initial parameters considered for any cycle $J_x=J_c=0, J_b=J$, these eigenstates in Eq.~\eqref{eq:cnot_states_app} reduce to the two qubits encoding states $|s,p\ra$ for  $p=0,1$ and  $s = 0,1$
\begin{eqnarray}
    |\psi_{1}^\text{c}(0)\ra&=& \frac{|c_0\ra-|d_0\ra}{\sqrt{2}} \equiv |0,0\ra  \\ \nonumber
    |\psi_{2}^\text{c}(0)\ra&=& \frac{|c_1\ra-|d_1\ra}{\sqrt{2}} \equiv |0,1\ra  \\ \nonumber
    |\psi_{3}^\text{c}(0)\ra&=& \frac{|c_0\ra+|d_0\ra}{\sqrt{2}} \equiv |1,0\ra  \\ \nonumber
    |\psi_{4}^\text{c}(0)\ra&=& \frac{|c_1\ra+|d_1\ra}{\sqrt{2}} \equiv |1,1\ra
\label{eq:cnot_logic_states_app}
\end{eqnarray} 
We calculate each element of $\Gamma_{z}$ in Eq.~\eqref{eq:WZ_app}
\begin{equation}
[\Gamma_{z}]_{\nu\nu'}= {\bf A}_{\nu\nu'} \cdot {\bf V}
\label{eq:G_gen_app}
\end{equation}
where  ${\bf V}=(\dot{J}_b,\dot{J}_c,\dot{J}_x)$ and ${\bf A}_{\nu\nu'}$ are given by
\begin{equation}
\label{eq:Aaux_app}
\begin{split}
A_{\nu\nu'} &= i \left\{  \langle \psi_{\nu} | \partial_{J_b} |\psi_{\nu'} \rangle, \langle \psi_{\nu}| \partial_{J_c} | \psi_{\nu'} \rangle , %\right. \\ &\left.\qquad 
\langle \psi_{\nu}|\partial_{J_x}| \psi_{\nu'} \rangle \right\} 
\end{split}
\end{equation}
Since all eigenvectors are real, then $\langle \psi_{\nu} | \partial_{J_\ell} |\psi_{\nu} \rangle = 0$ for all each $\nu=1,\dots,4$ and for each $\ell=b,c,x$ . Hence, the diagonal elements of the connection $\Gamma_z$ are zero. Furthermore, since $|\psi_{1}\ra$ and $|\psi_{2}\ra$ and independent on the parameter $J_\ell$, the only non-zero terms in $\Gamma_z$ are those induced by $|\psi_{3}\ra$ and  $|\psi_{4}\ra$. 
It then follows %for  $|\psi_{sL}\ra$ and  $|\psi_{sR}\ra$
\begin{equation}
\footnotesize
A_{34} 
%=   \frac{i}{( 2J_c^2+ J_x^2 + J_b^2 ) \sqrt{2J_c^2+(J_x + J_b)^2}\sqrt{2J_c^2+(J_x - J_b)^2}} 
=   \frac{i}{R_3^2 N_+ N_-} 
\left\{ J_x (J_b^2 - 2 J_c^2 - J_x^2) , 4  J_b J_c J_x , J_b ( J_x^2-J_b^2 - 2 J_c^2) \right\}   
\label{eq:Gaux_app}
\end{equation}
with $A_{43}  = -A_{34}$. The Wilczek-Zee connection then reads
\begin{equation}
\begin{split}
%\footnotesize
\Gamma_z 
&=  \frac{ 2 J_x J_c^2  \dot{J}_b -4 J_b J_cJ_x \dot{J}_c +2 J_b J_c^2 \dot{J}_x  }{R_3^2 N_+ N_-}  \\
 &  \frac{- J_b J_x (J_b \dot{J}_b  +  J_x \dot{J}_x)  +  J_x^3 \dot{J}_b + J_b^3 \dot{J}_x   }{R_3^2 N_+ N_-} 
   \frac{\sigma_0 - \sigma_z}{2}\otimes\sigma_y  \\
   & = f_\text{CN}(J_c,J_b,J_x) (\sigma_0 - \sigma_z)\otimes\sigma_y
\end{split}  
\label{eq:G_2Q_app}
\end{equation}
for the Pauli matrices $\sigma_p$ with $i=,x,y,z$ and the identity $\sigma_0$. 
For a closed path $\gamma$ in the parameter space $\{J_b,J_c,J_x\}$, let us define the result of the path integral yields $\eta=\oint_\gamma [\Gamma_z ]_{34} d\gamma$. In Eq.~\eqref{eq:holo_app}, this results 
\begin{equation}
\small
\begin{split}
W(z_0,z_1) &=  \frac{\sigma_0 - \sigma_z}{2}\otimes\left[ \cos\eta \sigma_0 + i \sin\eta \sigma_y  \right] + \frac{\sigma_0 + \sigma_z}{2}\otimes \sigma_0 \\
&= 
   \begin{pmatrix} 
   1&0&0&0 \\
   0&1&0&0 \\
   0&0&\cos\eta&\sin\eta \\
   0&0&-\sin\eta&\cos\eta 
   \end{pmatrix} 
\end{split}
\label{eq:holo_angle}
\end{equation}
Let us consider the driving cycle $\gamma_{\text{C}}$ for the ${\rm CNOT}$, and split it in three separate branches $\ell_1$, $\ell_2$ and $\ell_3$ respectively colored in red, blue and green. 
\begin{figure}[htbp]
	\centering
	\includegraphics[width=0.45\columnwidth]{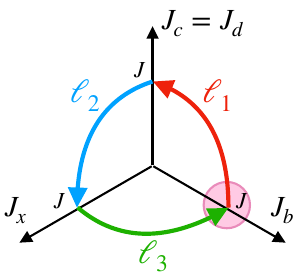}  
	\caption{Driving cycle $\gamma_{\text{C}}$ split in three branches $\ell_1$ (red), $\ell_2$ (blue) and $\ell_3$ (green). }
	\label{fig:cycle_app}
\end{figure} 
In each branch $\ell_1$, $\ell_2$ and $\ell_3$, only pairs of parameters evolve while the third is set to zero. This immediately sets the first three terms in Eq.~\eqref{eq:G_2Q_app} irrelevant as they require the simultaneous variations of all three parameters.  
The four remaining terms require the simultaneous variation of $J_b$ and $J_x$ -- hence, the branches $\ell_1$ and $\ell_2$ immediately yield zero. 
Branch $\ell_3$ instead reads set $J_b = \cos\theta$ and $J_x = \sin\theta$, with $J_c \equiv 0$. 
From Eq.~\eqref{eq:G_2Q_app} 
\begin{equation}
\begin{split}
[\Gamma_z]_{34} 
&=  \frac{  J_x^3 \dot{J}_b + J_b^3 \dot{J}_x - J_b J_x (J_b \dot{J}_b  +  J_x \dot{J}_x) }{( J_x^2 + J_b^2 )  (J_x^2 - J_b^2) }  \\
&=  \frac{ \cos^4\theta  -  \sin^4 \theta   }{\sin^2\theta - \cos^2\theta } = 1  \\
%&=  \frac{ (\cos^2\theta  -  \sin^2 \theta)(\cos^2\theta  +  \sin^2 \theta   )   }{\sin^2\theta - \cos^2\theta }  = 1 
\end{split}  
\label{eq:G_cycle}
\end{equation}
Integrated from $0$ to $\frac{\pi}{2}$, it follows that $\eta = \int_0^{\frac{\pi}{2}}  [\Gamma_z]_{34} dz  = \frac{\pi}{2}$ which, via Eq.~\eqref{eq:holo_angle} implies that the holonomy $W$ is 
\begin{equation}
\begin{split}
&W_\text{CN}= 
\begin{pmatrix} 
   1&0&0&0 \\
   0&1&0&0 \\
   0&0&0&1 \\
   0&0&-1&0 
   \end{pmatrix}  \\
 = &  \frac{1}{2}[(\sigma_0+\sigma_z)\otimes\sigma_0+i(\sigma_0-\sigma_z)\otimes \sigma_y]
\end{split}  
\label{eq:holo_cnot_app}
\end{equation}

\noindent
For the ${\rm SWAP}$, we keep the symmetry conditions $J_{x_L}=J_{x_R}\equiv J_{x}$, $J_{b_L}=J_{b_R}\equiv J_{b}$ and $J_{c_L}=J_{d_L}\equiv J_{c}$ set for the ${\rm CNOT}$, and impose an anti-symmetry condition $J_{c_R}= - J_{d_R}\equiv J_{c}$. 
The orthonormal basis of zero-modes  for this specific driving cycle -- hence, we denote the basis with an apex s 
\begin{eqnarray}
  |\psi_{1}^\text{s}(z)\ra&=&  \frac{|c_0\ra-|d_0\ra}{\sqrt{2}} \\ \nonumber
     |\psi_{2}^\text{s}(z)\ra&=&
   \frac{ J_x |s_0\ra +  J_b |a_1\ra - \sqrt{2}J_c |b_1\ra }{R_3} \\ \nonumber
   |\psi_{3}^\text{s}(z)\ra&=&     \frac{J_b ( 2J_c^2 -  J_x^2 +  J_b^2 )  |s_0\ra +
J_x (2J_c^2  +  J_x^2 -  J_b^2 )  |a_1\ra}{R_4} \\
&+&    \frac{\sqrt{2} J_c (J_b^2 + 2 J_c^2 + J_x^2) |b_0\ra -  2 \sqrt{2} J_x J_b J_c  |b_1\ra }{R_3} \\ 
   |\psi_{4}^\text{s}(z)\ra&=& \frac{|c_1\ra+|d_1\ra}{\sqrt{2}} 
\nonumber
\label{eq:swap_states_app}
\end{eqnarray} 
where $|a_1\ra= \frac{|c_1\ra- |d_1\ra}{\sqrt{2}}$. 
Alike for the ${\rm CNOT}$, at $J_x=J_c=0, J_b=J$ these eigenstates in Eq.~\eqref{eq:swap_states_app} reduce to the two qubits encoding states $|s,p\ra$ for  $p=0,1$ and  $s = 0,1$ as in Eq.~\eqref{eq:cnot_logic_states_app}. 
For the same driving cycle in Fig.~\ref{fig:cycle_app}, the holonomy results in
\begin{equation}
\begin{split}
&W_\text{SW}= 
   \begin{pmatrix} 
   1&0&0&0 \\
   0&0&1&0 \\
   0&-1&0&0 \\
   0&0&0&1 
   \end{pmatrix} \\
   = & \frac{1}{2}(\sigma_0\otimes\sigma_0-i\sigma_x\otimes \sigma_y+i\sigma_y\otimes \sigma_x+\sigma_z\otimes \sigma_z)
\end{split}  
\label{eq:holo_swap_app}
\end{equation}

\noindent
Let us obtain the ${\rm SWAP}$ from the ${\rm CNOT}$. 
An holonomic NOT gate acting on the control qubit on the right tripod $H_R$ can be obtained in Eq.~\eqref{eq:holo_sigmaY_angle} by setting $\eta_L=0$ and $\eta_R\frac{\pi}{2}$. These two angles are obtained by (i) in the left tripod $H_L$ keep all the hopping to zero $J_{\ell_L} = 0$ for $\ell = c,d,b$, while (ii) running twice the driving cycle in Fig.~\eqref{fig:cycle_sigmaY_app} in the right tripod $H_R$. Eq.~\eqref{eq:holo_sigmaY_angle} then reduces to
\begin{equation}
\begin{split}
&W_\text{nR} = 
\begin{pmatrix} 
   1&0&0&0 \\
   0&0 &0&1 \\
   0&0&1&0 \\
   0&-1&0&0 
   \end{pmatrix} \\
   = &  \sigma_0  \otimes \frac{\sigma_0 + \sigma_z}{2} + i \sigma_y \otimes \frac{\sigma_0 - \sigma_z}{2} 
\end{split}
\label{eq:holo_sigmaY_L0_Rpi2}
\end{equation}

\noindent
The product 
\begin{equation}
\begin{split}
 W_\text{nR}^\dagger W_\text{CN} W_\text{nR} = 
   \begin{pmatrix} 
    1&0&0&0 \\
   0&0&1&0 \\
   0&-1&0&0 \\
   0&0&0&1
   \end{pmatrix}
= W_\text{SW}
\end{split}  
\label{eq:holo_swap_v2_app}
\end{equation}
corresponds to the holonomy $W_\text{SW}$ of the ${\rm SWAP}$ gate in Eq.~\eqref{eq:holo_swap_app}.

\section{Deutsch's algorithm}\label{app:DeuAlg}

The Deustch's algorithm leverages on quantum superposition to distinguish binary functions $f$ in Table~\ref{tab:1} between constant and balanced with a single measurement.  
Formally, for a binary function $f$ and a state $|s,p\ra$ with control qubit $s$ and target qubit $p$, the oracle $U_f$ of the Deustch's algorithm acts as 
$U_f(|s,p\ra) = |s,p\oplus f(s)\ra$  the XOR $a\oplus b = [a+b]_2$ (sum modulo 2).

Applied on a superposition
\be
|s,-\ra = \frac{1}{\sqrt{2}} (|s,0\ra - |s,1\ra)
\label{eq:SP_state0_app}
\ee 
the oracle $U_f$ yields the {\it phase kickback} property $U_f|s,-\ra=(-1)^{f(s)}|s,-\ra$. 
For a superposition state 
\be 
\begin{split}
    |+,-\ra &= \frac{1}{\sqrt{2}} (|0,-\ra + |1,-\ra)\\
    &= \frac{1}{2} (|0,0\ra - |0,1\ra + |1,0\ra - |1,1\ra)
\end{split} 
\label{eq:SP_state_app}
\ee
obtained by applying the Hadamard gates on both control and target on an initial input $|0,1\ra$, it then follows that 
\begin{equation}
\begin{split}
   U_f|+,-\ra &=|-\ra\frac{(-1)^{f(0)}|0\ra+(-1)^{f(1)}|1\ra}{\sqrt{2}} =
   \\
   &=|-\ra(-1)^{f(0)}\frac{|0\ra+(-1)^{f(0)\otimes f(1)}|1\ra}{\sqrt{2}}=
   \\
 &=\left\{\begin{array}{@{}l@{}}
     (-1)^{f(0)}|+,-\ra \qquad f(0)\otimes f(1)=0\\
       (-1)^{f(0)}|-,-\ra \qquad f(0)\otimes f(1)=1
   \end{array}\right.\,
   \end{split}
   \label{eq:deu_funct}
 \end{equation}
\\
A final Hadamard gate on the control qubit $s$ allows then to distinguish constant functions from balanced ones.

We now explicitly show that the photonic array proposed in Fig.~\ref{fig:1}(a) and the driving cycle in Fig.~\ref{fig:2}(b)  with proper signs of the hopping terms implement the oracle identity in Eq.~\eqref{eq:deu_funct}. 
Consider the Hamiltonian  $H =  H_{0}  + H_{1} + H_{T}$, where  $H_{T} = J_{x_0} b_1^\dagger u_0 + J_{x_1} b_0^\dagger u_1 + \text{H.c.}$ and for $p=0,1$  we have that $H_p = (J_{b_p} b_p^\dagger + J_{c_p} c_p^\dagger + \alpha_p J_{c_p} d_p^\dagger) u_p + \text{H.c.}$ with $J_{d_p} = \alpha_p J_{c_p}$ for $\alpha_p=\pm 1$. 
The driving cycle in Fig.~\ref{fig:2}(b) yields four holonomies $W_{\alpha_0,\alpha_1}$ 
\begin{equation}
\begin{split}
  W_{1,1}= &\frac{1}{2}[(\sigma_0+\sigma_z)\otimes\sigma_0+i(\sigma_0-\sigma_z)\otimes \sigma_y]
  \\
  W_{1,-1}= &\frac{1}{2}[\sigma_0\otimes \sigma_0+\sigma_z\otimes \sigma_z+i(\sigma_y\otimes \sigma_x-\sigma_x\otimes \sigma_y)]
    \\
W_{-1,1}= &\frac{1}{2}[\sigma_0\otimes \sigma_0-\sigma_z\otimes \sigma_z+i(\sigma_y\otimes \sigma_x+\sigma_x\otimes \sigma_y)]
\\
W_{-1,-1}= &\frac{1}{2}[(\sigma_0-\sigma_z)\otimes\sigma_0+i(\sigma_0+\sigma_z)\otimes \sigma_y]
\end{split}
\label{}
\end{equation}
Straightening the minus signs in the holonomies via the phase gate, we obtain the four transformations 
\begin{equation}
\begin{split}
V_{f_1}&= \begin{pmatrix} 
   1&0&0&0 \\
   0&1&0&0 \\
   0&0&0&1 \\
   0&0&1&0 
   \end{pmatrix}   
   \quad
V_{f_4}= \begin{pmatrix} 
   0&1&0&0 \\
   1&0&0&0 \\
   0&0&1&0 \\
   0&0&0&1 
   \end{pmatrix}  
   \quad 
   \\  
V_{f_2}&= \begin{pmatrix} 
   1&0&0&0 \\
   0&0&1&0 \\
   0&1&0&0 \\
   0&0&0&1 
   \end{pmatrix}   
\quad
V_{f_3}= \begin{pmatrix} 
   0&0&0&1 \\
   0&1&0&0 \\
   0&0&1&0 \\
   1&0&0&0 
   \end{pmatrix}      
\end{split}  
\label{eq:S_ora_const_us}
\end{equation}
We define the functions $f_1,\dots,f_4$ in Table~\ref{tab:1} via the transformations $V_{f_1},\dots,V_{f_4}$ in Eq.~\eqref{eq:S_ora_const_us} by firstly setting the input $x\in\{0,1\}$ in the two-qubits states as 
\be
\begin{split}
\text{input }\,\mathcal{I}\,:
\begin{cases}
x=0\,\,\longmapsto\,\, |0,0\ra \\
x=1\,\,\longmapsto\,\, |1,1\ra 
\end{cases}
\end{split}
\label{eq:S_end_p0_us}
\ee
Then we compose $V_{f_1},\dots,V_{f_4}$ with a measurement done via scalar product $P_1 = \la1,0|+\la1 ,1|$ 
\be
\begin{split}
f_\ell(x) = 
|P_1\circ V_{f_\ell} |x,x\ra |^2 
\end{split}
\label{eq:f1_f4_us}
\ee
Indeed, 
\be
\begin{split}
f_1(x) &= 
\begin{cases}
0 \xrightarrow{\mathcal{I}} |0,0\ra \xrightarrow{V_{f_1}} |0,0\ra \xrightarrow{P_1} 0 \\
1 \xrightarrow{\mathcal{I}}|1,1\ra \xrightarrow{V_{f_1}} |1,0\ra \xrightarrow{P_1} 0 
\end{cases}\\
f_2(x) &= 
\begin{cases}
0 \xrightarrow{\mathcal{I}} |0,0\ra \xrightarrow{V_{f_2}} |0,0\ra \xrightarrow{P_1} 0 \\
1 \xrightarrow{\mathcal{I}} |1,1\ra \xrightarrow{V_{f_2}} |1,1\ra \xrightarrow{P_1} 1 
\end{cases}\\
f_3(x) &= 
\begin{cases}
0 \xrightarrow{\mathcal{I}} |0,0\ra \xrightarrow{V_{f_3}} |0,1\ra \xrightarrow{P_1} 1 \\
1 \xrightarrow{\mathcal{I}}|1,1\ra \xrightarrow{V_{f_3}} |1,0\ra \xrightarrow{P_1} 0 
\end{cases}\\
f_4(x) &= 
\begin{cases}
0 \xrightarrow{\mathcal{I}} |0,0\ra \xrightarrow{V_{f_4}} |0,1\ra \xrightarrow{P_1} 1 \\
1 \xrightarrow{\mathcal{I}}|1,1\ra \xrightarrow{V_{f_4}} |1,1\ra \xrightarrow{P_1} 1 
\end{cases}\\
\end{split}
\label{eq:f1_f4_us2}
\ee
The superposition states where we perform the Deustch's algorithm are 
 \be
\begin{split}
|+,-\ra %&= \frac{1}{\sqrt{2}} (|0,-\ra + |1,-\ra) \\
&= \frac{1}{2} (|0,0\ra - |0,1\ra + |1,0\ra-|1,1\ra) \\
|-,-\ra %&= \frac{1}{\sqrt{2}} (|0,+\ra + |1,+\ra) \\
&= \frac{1}{2} (|0,0\ra - |0,1\ra - |1,0\ra + |1,1\ra) 
\end{split}
\label{eq:state_pm}
\ee
Then 
\be
\small
\begin{split} 
   V_{f_1}(|-,-\ra) %&= \frac{1}{2} U_{1,1}  (|0,0\ra - |0,1\ra - |1,0\ra + |1,1\ra ) \\
   &= \frac{1}{2} (|0,0\ra - |0,1\ra - |1,1\ra + |1,0\ra ) = |+,-\ra
     \\
   V_{f_2}(|-,-\ra) %&= \frac{1}{2} U_{1,-1}  (|0,0\ra - |0,1\ra - |1,0\ra + |1,1\ra) \\
   &= \frac{1}{2}  (|0,0\ra - |1,0\ra - |0,1\ra + |1,1\ra ) = |-,-\ra
    \\
   V_{f_3}(|-,-\ra) %&= \frac{1}{2} U_{-1,1}  (|0,0\ra - |0,1\ra - |1,0\ra + |1,1\ra)  \\
   &= \frac{1}{2}   (|1,1\ra - |0,1\ra - |1,0\ra + |0,0\ra) = -   |-,-\ra
   \\    
   V_{f_4}(|-,-\ra) %&= \frac{1}{2} U_{-1,-1}  (|0,0\ra - |0,1\ra - |1,0\ra + |1,1\ra)  \\
   &= \frac{1}{2}  (|0,1\ra - |0,0\ra - |1,0\ra + |1,1\ra) = - |+,-\ra
   \end{split}
\label{eq:Ws1s2}
\ee

\bibliography{Qgates}

\end{document}